\documentclass[aps,pre,reprint]{revtex4-1}

\usepackage[english]{babel}
\usepackage[latin1]{inputenc}

\usepackage{graphics}
\usepackage{graphicx}
\usepackage{xcolor}

\usepackage{amsfonts}
\usepackage{amssymb}

\definecolor{Pergamen}{RGB}{235,225,200}
\definecolor{LightGray}{RGB}{235,235,230}
\definecolor{PaleBlue}{RGB}{190,210,255}

\definecolor{Red}{RGB}{255,0,0}

\usepackage{setspace}
\usepackage{textpos}

\usepackage{amsmath}

\begin{document}
\title{%
	{\bf Dynamical Stationarity as a Result of Sustained Random Growth}
}


\author{Tam\'as~S.~ Bir\'o\thanks{Biro.Tamas@wigner.mta.hu}}
\affiliation{%
	MTA Wigner Research Centre for Physics, Budapest, Hungary
}
\author{Zolt\'an~N\'eda\thanks{zneda@ubbcluj.ro}}
\affiliation{%
	Babes-Bolyai University, Department of Physics, Cluj, Romania
}

\date{\today}


\newcommand{\vs}{\vspace{3mm}}

\newcommand{\be}{\begin{equation}}
\newcommand{\ee}[1]{\label{#1} \end{equation}}
\newcommand{\ba}{\begin{eqnarray}}
\newcommand{\ea}[1]{\label{#1} \end{eqnarray}}
\newcommand{\nl}{\nonumber \\}
\newcommand{\re}[1]{(\ref{#1})}
\newcommand{\spr}[2]{\vec{#1}\cdot\vec{#2}}
\newcommand{\ave}{\overline{u}}
\newcommand{\ve}[1]{\left\vert #1  \right\vert}
\newcommand{\exv}[1]{ \left\langle {#1} \right\rangle}

\newcommand{\pd}[2]{ \frac{\partial #1}{\partial #2}}
\newcommand{\pt}[2]{ \frac{{\rm d} #1}{{\rm d} #2}}
\newcommand{\pv}[2]{ \frac{\delta #1}{\delta #2}}

\newcommand{\grad}{{\vec{\nabla}}}

\newcommand{\ead}[1]{ {\rm e}^{#1}}
\newcommand{\infi}{ \int_0^{\infty}\limits\!}
\newcommand{\sumi}{ \sum_{n=0}^{\infty}\limits\!}


\begin{abstract}

 In sustained growth with random dynamics stationary distributions can exist without detailed balance.
 This suggests thermodynamical behavior in fast growing complex systems.
 In order to model such phenomena we apply both a discrete and a continuous master equation. 
 The derivation of elementary rates from known stationary distributions
 is a generalization of the fluctuation--dissipation theorem.
 Entropic distance evolution is given for such systems.
 We reconstruct distributions obtained for growing networks, particle production,
 scientific citations and income distribution.

\end{abstract}

\maketitle


\section{Introduction}

Statistical physics methods are applied to problems related to complex system evolution
in an increasing manner. While these are powerful enough to describe essential
properties of statistical data and their distributions, the meaning of parameters
behind such distributions can be understood deeper if derived from dynamical models.
Following Occam's razor principle 
(among competing hypotheses, the one with the fewest assumptions should be selected),
simple rules for the dynamics are welcome.

The dynamics of many complex systems can be studied by using a simple master-equation 
approach \cite{MASTERPAPER,MODCOMP,COHEN,GUIDE}. 
Beside physics, such studies are also popular in network science 
\cite{BARABASI-PHYSICA,BARABASI-REVMOD,DOROG1,DOROG2},
biology \cite{MASTERBIO}, economics \cite{STANLEY}, chemistry \cite{VANKAMPEN}, 
epidemics \cite{EPIDEMI,EPIBOOK}, scientometrics \cite{SCHUBERT,EVOLSOCNET} and 
sociology \cite{SOCIO}. 
Generally such dynamical processes tend to a stationary state with an 
invariant limiting distribution \cite{MATHBOOK}.

In the master equation approach to the evolution of general probability distributions,
we know several statements for systems satisfying the detailed balance condition
in their stationary state \cite{BIRONEDA}, but much less is known for fast growing complex systems
without detailed balance. In particular, if the microprocesses are not reversible,
the entropy growth and the global stability of stationary solutions are not guaranteed
even for generalized entropies. Such cases occur in open systems.

In this paper we investigate a promising subset of unbalanced master equations leading to stationary
distributions. Such an approach can be applied to understand several 
complex phenomena. In this work we refer to application examples
for emerging particle distributions in high-energy accelerator experiments, to 
income distributions following from redistribution and taxation strategies,
to scinetific citation dynamics and to evolution of growing complex networks.

In this framework, the stationary distribution, $Q_n$, is determined by two microscopic
rates: $\mu_n$ describes the transition rate from a state with $n$ quanta to $n+1$ inside a chain of
states, while $\gamma_n$ describes a loss rate for the state $n$ towards an unspecified
environment. We assume that there is no $n$ to $n-1$ process, so the transition dynamics is unidirectional.
Without having a state dependent loss rate, $\gamma_n=0$, the only
possibility for a stationary distribution would be $Q_n \sim 1/\mu_n$, a trivial case.
Already for constant and linearly $n$-dependent rates a rich structure of possible solutions emerges.

\begin{figure}[h]
\centerline{\includegraphics[width=0.25\textwidth]{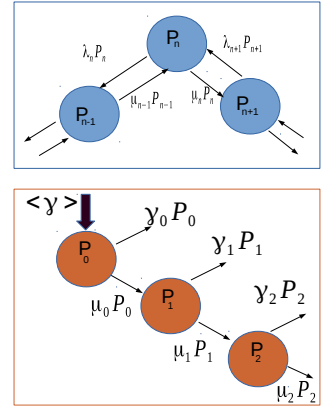} }
\caption{~\label{PICTOR}
Schematic view of master equations for balanced (top) 
and sustained random growth (bottom) processes.
}
\end{figure}

Since there is no reverse process inside the chain of states, a detailed balance
condition cannot be fulfilled. We illustrate the difference between the classical
scheme allowing detailed balance and the presently discussed one-sided growth
picture with the flow diagrams on Fig.\ref{PICTOR}. 

State-dependent loss rates, $\gamma_n \ne 0$, open the door to nontrivial stationary distributions.
Usually models are constructed with assumed transition rates $\mu_n$ and $\gamma_n$
and the stationary (limiting) distribution, $Q_n$, is derived. However, the reverse
problem is also interesting: by observing a distribution, $Q_n$, and knowing
the interaction rate with the environment, $\gamma_n$, one wishes to reconstruct
the internal dynamics of the system governed by the  rates, $\mu_n$.
We interpret the quantity $n$ in high-energy experiments as number of hadrons
produced in energetic collisions. In studying income distributions
$n$ is the amount of money received by consumer units. 
In studying the impact of scientific papers $n$ is the number of received citations.
For complex random networks $n$ means the number of connections
starting from a given node and $Q_n$ is therefore the degree distribution.

We present both a master equation approach over discrete states labelled by $n$,
and its continuous limit. Finally the stability
of stationary distributions obtained from given transition rates is investigated
in terms of a generalized entropic distance.

\section{Master and flow equation framework}


Now we turn to the definition of the underlying mathematical formalism.
We consider linear and first order time evolution equations for the distribution,
$P_n(t)$ and its continous version, $P(x,t)$. The corresponding stationary
distributions, $Q_n$ and $Q(x)$, respectively, shall be determined by the same
equations with vanishing time derivative. Beyond finding out what stationary
distributions, i.e. results of the long term evolution, belong to given rates
$\mu_n, \gamma_n$ (or $\mu(x)$, $\gamma(x)$) one is interested in the whole
process starting from arbitrary initial distributions as well as in
the stability and basin of attraction for the final distribution.


\subsection{Discrete state space master equation}

The sustained growth master equation hereafter is
given as depicted in the lower part of Fig.\ref{PICTOR}:
\be
 \dot{P}_n \: = \: \mu_{n-1}P_{n-1} \, - \, \left( \mu_n + \gamma_n \right) P_n
\ee{GROWPN}
for $n \ge 1$. 
The corresponding equation for the $n=0$ term can be obtained 
from the normalization condition $\sum_{n=0}^{\infty}\limits P_n(t)=1$:
\be
 \dot{P}_0 \: = \: \exv{\gamma}_P - \left( \mu_0 + \gamma_0 \right) P_0.
\ee{GROWZERO}
Here we used the abbreviation $\exv{\gamma}_P = \sum_{n=0}^{\infty}\limits \gamma_n P_n$.
This system allows for stationary solutions satisfying:
\be
 \mu_{n-1}Q_{n-1} \: = \: (\mu_n + \gamma_n) Q_n
\ee{GROWSTAT}
for $n\ge 1$ and $Q_0 = \exv{\gamma}_Q / (\mu_0 + \gamma_0)$. 
Eqs. (\ref{GROWPN}) and (\ref{GROWZERO}) constitute a specific realization of a general, 
continuous-time Markov process:
\be
 \dot{P}_n \: = \: \sum_m\limits 
 \left( w_{n\leftarrow m} P_m - w_{m \leftarrow n} P_n \right)
\ee{GENMARKOV}
with
\be
 w_{n\leftarrow m} \: = \: \mu_m \delta_{m,n-1} \, + \, \gamma_m \delta_{n,0}.
\ee{SPECMARKOV}
The inflow and outflow in each patch in Fig.\ref{PICTOR} balance each other
in the stationary state.
This also offers a strategy to
reconstruct the link connection probability rate to a link with already
$m$ connections or to increase a conveniently discretized income from $m$ to $m+1$, $\mu_m$, by
observing the stationary distribution, $Q_n$, and the loss rate $\gamma_n$.
We simply sum up eq.(\ref{GROWSTAT}) from $n=m+1$ to infinity and obtain
\be
\mu_m \: = \:  \frac{1}{Q_m} \sum_{n=m+1}^{\infty}\limits \gamma_n \, Q_n.
\ee{MUZERO}
This relation reminds to the fluctuation--dissipation theorem,
in particular when the stationary distribution is exponential, 
$Q_n=\ead{-\beta n}/Z=(1-q)q^n$, and the loss rate
due to environmental effects is constant $\gamma_n=\gamma$. In this case
eq.(\ref{MUZERO}) delivers a constant inner-chain rate reminding to the quantum Kubo formula:
\be
 \mu_m^{{\rm exp }} \: = \: \gamma \, \frac{q}{1-q} \: = \: \gamma \, \frac{1}{\ead{\beta}-1}.
\ee{MUZERO_SPECIAL}
The general solution of the recursion represented by eq.(\ref{GROWSTAT}) is given as a ratio of
$n$-fold products,
\be
 Q_n \: = \: Q_0 \, \frac{\prod_{i=0}^{n-1}\limits \mu_i}{\prod_{j=1}^n\limits (\mu_j+\gamma_j)}. 
\ee{QNGR}
$Q_0$ can either be obtained from the normalization condition $\sumi Q_n = 1$ or
applying eq.(\ref{GROWZERO}) with $\dot{Q}_0=0$. It is not trivial that these are equivalent
procedures: the product form (\ref{QNGR}) and the definition of the expectation value
delivers
\be
 Q_0 = \frac{\exv{\gamma}_{Q}}{\mu_0+\gamma_0} \: = \:  
 Q_0 \, \sum_{n=0}^{\infty}\limits \frac{\gamma_n}{\mu_n} \prod_{i=0}^n\limits 
 \frac{\mu_i}{\mu_i+\gamma_i}.
\ee{CONSEQUENTGAMMA}
Consistency can easily be re-formulated in terms of the basic ratios, $r_i=\gamma_i/\mu_i$,
after dividing both sides with $Q_0 \ne 0$:
\be
 1 \: = \: \sum_{n=0}^{\infty}\limits r_n \cdot \prod_{i=0}^n\limits \frac{1}{1+r_i}.
\ee{CONSISTENT}
It is at the first glance surprising, but true, that this identity is fulfilled for 
any infinite series of $r_i\ne -1$ ratios. A short mathematical proof is given in the 
Appendix.

\subsection{Continuum approach}

It is instructive to obtain the above equations in a continuous version.
We set up the following Markovian framework:
\be
 \pd{}{t} P(x,t) \: = \: \int \left[w(x,y) P(y) - w(y,x) P(x) \right] \, dy
\ee{GENCONT}
with
\be
 w(x,y) \: = \: \frac{1}{\Delta x} \mu(y) \, \delta(y-x+\Delta x) \, + \, \gamma(y) \, \delta(x).
\ee{CONTWXY}
Next we take the $\Delta x \to 0$ limit leading to
\be
\pd{}{t} P(x,t) \: = \: -\pd{}{x} \left( \mu(x) P(x,t) \right) - \gamma(x) P(x,t) + \exv{\gamma}_P \delta(x).
\ee{GROWCONT}
Please note that this is an integro-differential equation containing
\be
 \exv{\gamma}_P \: = \: \int\! \gamma(y) P(y,t) \, dy.
\ee{GAMMAWITHP}
Equation (\ref{GROWCONT}) desribes a flow with general velocity field, $\mu(x)$, a loss  rate $\gamma(x)$
and a feeding at $x=0$.
From now on we restrict the discussion to $x>0$, and all effects stemming from the singular
term $\exv{\gamma}_P\delta(x)$ are treated by enforcing the normalization condition.

The stationary distribution in the continuous model satisfies
\be
\pt{}{x} \left( \mu(x) Q(x) \right) \: = \: -\gamma(x) \, Q(x),
\ee{CONTSTAT}
revealing the solution
\be
Q(x) \: = \:  \frac{K}{\mu(x)} \, \ead{-\displaystyle\int \!\frac{\gamma(x)}{\mu(x)} \, dx}.
\ee{CONTSOLUT}
The constant, $K$, is specified by the normalization $\infi Q(x) dx = 1$.

We note that this form can also directly be obtained from the discrete solution, 
eq.(\ref{QNGR}), when it is written in the alternative form
\be
Q_n \: = \: Q_0 \frac{\mu_0}{\mu_n} \, \ead{-\sum_{j=1}^n\limits 
 \ln\left(1+\frac{\gamma_j}{\mu_j}\right)}.
\ee{QNGRMOD}
The approximation, $\gamma_j/\mu_j = \gamma(x) \Delta x/\mu(x)  \ll 1$,
defines the continuous limit and one arrives at
\be
Q_n \: \approx \: Q_0 \frac{\mu_0}{\mu_n} \, 
 \ead{- \sum_{j=1}^n\limits \frac{\gamma(j\Delta x)}{\mu(j\Delta x)} \, \Delta x },
\ee{QNEXPSUM}
to an obvious analog of eq.(\ref{CONTSOLUT}) with $K=\mu(0)Q(0)\Delta x$.

The inner-chain growth rate, $\mu(x)$, can be reconstructed
from the known stationary distribution, $Q(x)$, and loss rate, $\gamma(x)$:
\be
 \mu(x) \: = \: \frac{1}{Q(x)} \, \int_x^{\infty}\limits \!\gamma(u) Q(u) \, du.
\ee{FLUDICONT}
The validity of this formula is tested by applying a derivation with respect to $x$
and eq.(\ref{CONTSTAT}).
For the exponential distribution, $Q(x) \sim \ead{-x/T}$, and constant $\gamma(x)=\gamma$ we obtain
\be
 \mu^{{\rm exp}}(x) \: = \: \gamma \, T.
\ee{FLUDISPECIAL}
The temperature-like parameter in the exponential distribution, $T$, becomes a factor between
two elementary rates $\gamma$ and $\mu$. In a physical picture $\gamma$ describes
dissipation, $\mu(x)$  random advances towards larger $x$ values.

\section{Particular rates and distributions}

In the followings we discuss the simplest choices for the involved rates.
First we keep the loss rate a positive constant, $\gamma_n=\gamma>0$, and vary the growth rate, $\mu_n$.
This is relevant for a wide class of distributions considered in statistics.
For a constant $\mu_j=\sigma$ we obtain the {\em geometrical distribution},
\be
Q_n \: = \: \frac{\gamma}{\sigma}\left(1+\gamma/\sigma\right)^{-1-n},
\ee{QNCONST}
shortly $Q_n=(1-q)q^n$, with $q=\sigma/(\sigma+\gamma)$. This is also called 
the {\em exponential}, or {\em Boltzmann--Gibbs distribution} in the form 
$Q_n=e^{-\beta n}/Z$ with $Z=1+\sigma/\gamma$ and $\beta = \ln(1+\gamma/\sigma)>0$.

For fast growing systems, like networks, citations or energetic hadronization,
the most prevalent is the next simplest case, $\mu_j=\sigma(j+b)$, describing
a growth rate with thresholded linear preference. Often $b=1$
is taken when investigating the evolution of degree distribution in networks \cite{KULLMAN-KERTESZ}.
Eq.(\ref{QNGR}) delivers
\be
Q_n \: = \: \frac{\gamma}{\sigma b + \gamma} \, \frac{(b)_n}{(b+1+\gamma/\sigma)_n}
\ee{QNLIN}
with the Pochhammer symbol: 
\be
(b)_n \: = \: b \cdot (b+1) \cdot \ldots \cdot (b+n-1) \: = \: \frac{\Gamma(b+n)}{\Gamma(b)}.
\ee{POCHHAMMER}
The {\em Waring distribution} \cite{IRWIN,GENERWARING,PREFAT} in eq.(\ref{QNLIN})
has a {\em power-law tail} for large $n$,
\be
 \lim_{n\to\infty}\limits Q_n \propto n^{-1-\gamma/\sigma}.
\ee{QNLARGE}
This behavior is based on the leading order behavior of Gamma functions for large arguments: 
\be
\lim_{n\to\infty}\limits n^{b-a} \frac{\Gamma(n+a)}{\Gamma(n+b)} \: = \: 1.
\ee{GAMMALIMIT}
Our result eq.\ref{QNLIN} coincides with eq.7. in Ref.\cite{KULLMAN-KERTESZ} at $b=1$.
The asymptotic power is steeper than minus one for positive rate factors $\gamma$ and $\sigma$, 
but it can be anything, depending on the
ratio of the universal driving rate $\gamma$ and the preference scale of the individual growth rates,
$\sigma = \mu_n-\mu_{n-1}$.
For $\gamma\to 0^+$ the stationary distribution tail (\ref{QNLARGE}) leads to {\em Zipf's law}
\cite{ZIPF}.

Now we analyze these particular growth rates by a constant loss rate, $\gamma(x)=\gamma$
in the continuous model.  We expect the same asymptotic behavior for the tail of the distribution.
For a constant growth rate, $\mu(x)=\sigma$, the stationary PDF becomes 
again the {\em exponential distribution},
\be
Q(x) \: = \: \frac{\gamma}{\sigma} \, \ead{- (\gamma/\sigma) \, x}.
\ee{CONTCONST}
This is the $\gamma \ll \sigma$ limit of the result in the discrete case, eq.(\ref{QNCONST}).
For a linearly preferential rate, $\mu(x)=\sigma\cdot(x+b)$, we obtain a {\em cut power-law}
in the form of the {\em Tsallis--Pareto distribution} 
\cite{THURNER,ZIPF,WIKI-PARETO,TSALLIS-ALB,DOROG4}
\be
Q(x) \: = \: \frac{\gamma}{b\sigma} \, \left(1+\frac{x}{b} \right)^{-1-\gamma/\sigma}.
\ee{CONTLINEARPREF}
Beyond these reassuring results a further question arises: 
what is the stationary distribution for weaker or stronger than
linear preferences in the attachment probability rate \cite{KRAPIVSKY-MASTER,KRAPIVSKY}? 
By assuming a general power,
$\mu(x)=\sigma(x+b)^a$, one obtains the {\em stretched exponential distribution},
\be
Q(x) \: = \: \frac{\gamma}{\sigma (x+b)^a} \, \ead{-\alpha(x+b)^{1-a}} 
\ee{QUADRATICMUN}
for $a<1$ with $\alpha = \gamma/[\sigma(1-a)]$. For $a > 1$ it  delivers 
$Q(x) \sim \gamma/\mu(x)$ tail. 
Eq.(\ref{QUADRATICMUN}) represents also a 3-parameter {\em Weibull distribution},
with $a=1-k$, $b=-\theta$ and $\gamma/\sigma = k\lambda^{-k}$ \cite{WIKI-WEIBULL}.
On the other hand for an exponential preference rate, $\mu(x)=\sigma \ead{\alpha x}$,
one obtains the {\em Gompertz distribution} \cite{GOMPERTZ}: 
\be
Q(x) \: = \: \frac{\gamma/\sigma}{1-\ead{-\gamma/\sigma\alpha}} \, \ead{-\alpha x} \, 
 \ead{-\frac{\gamma}{\alpha\sigma}(1-\ead{-\alpha x})}.
\ee{QXGOMPERTZ}
We note by passing that the particular form for the stationary
distribution, $Q(x)$ in eq.(\ref{CONTSOLUT}) with constant $\gamma$, 
is term by term compatible with the usual notions used in 
survival analysis in demography, finance and insurance statistics\cite{HAZARD}: 
the PDF (probability density function) has in this case the form
\be
 Q(x) \: = \: h(x) \, \ead{-H(x)}
\ee{PDFWITHAZARD}
with $h(x)$ being the {\em hazard rate} and $H(x)=\int_0^{x}\limits h(t) dt$ 
the {\em cumulative hazard}.
The factor $R(x)=\ead{-H(x)}$ is called {\em survival rate}.
The growth rate inside the chain is simply related to the hazard rate: $\mu(x)=\gamma/h(x)$.
This is again a special case for the fluctuation--dissipation correspondence
summarized in eq.(\ref{FLUDICONT}). 
The same relation has been called truncated expectation value theorem
in Refs.\cite{TELCS1,TELCS2}.
A similar result has been derived by generalizing the thermodynamical fluctuation--dissipation
relation between the diffusion and damping coefficients for a
general Fokker--Planck equation stemming from a particularly colored, i.e. energy dependent,
multiplicative noise Langevin equation \cite{BIRO-JAKOVAC,BIRO2006}
and eq.(5.46) in \cite{BIROBOOK}.

Finally we mention two important examples, frequently encountered
in complex system applications, which do not fit in the above scheme.
We consider loss rates, $\gamma_n$, which can be {\em negative} for some low $n$. 
Such a mechanism has been suggested a.o. in Ref.\cite{OSADA}
for describing the multiplicity distribution of hadrons in high-energy collision events.
The linear rates
\be
 \gamma_n \: = \: \sigma ( n - kf),  \qquad
 \mu_n \: = \: \sigma f (n+k)
\ee{NBDRATES}
will lead to a {\em negative binomial} stationary distribution: 
\be
 Q_n \: = \: Q_0 \frac{(f\sigma)^n \prod_{i=0}^{n-1}\limits (j+k)}{(\sigma(1+f))^n \prod_{i=1}^n\limits i} 
 \: = \: \binom{n+k-1}{n} \, f^n (1+f)^{-n-k}.
\ee{NBDFROMRATES}
We note that in this case $\exv{\gamma}_Q=0$. 
In order to achieve a normalized stationary distribution 
obviously $\gamma_n+\mu_n > 0$ for all $n$.

\begin{figure}[h]
\begin{center}
 \includegraphics[width=0.25\textwidth]{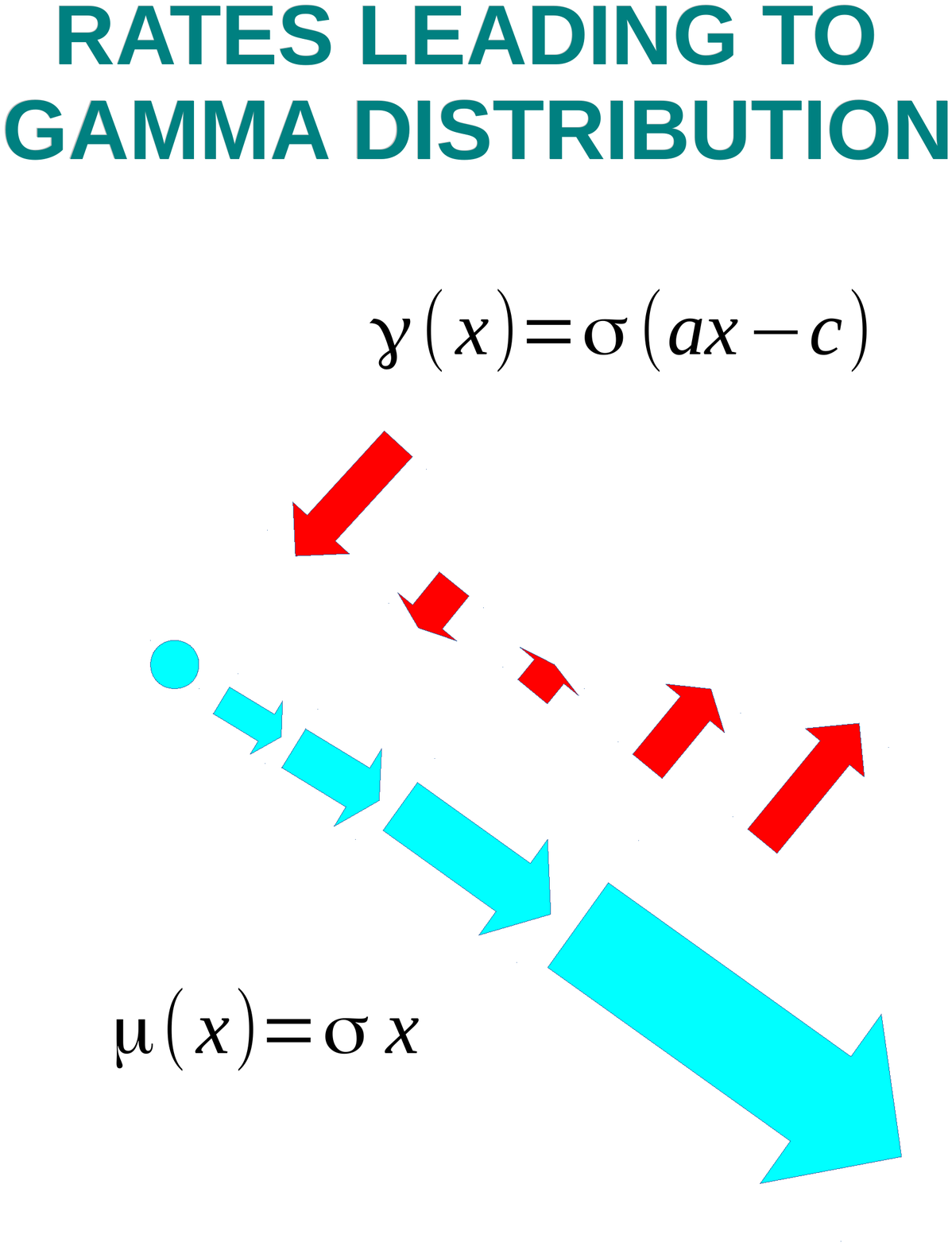}
\end{center}
\caption{\label{GAMMAFIG}
 Scheme of the linear rates leading to the gamma distribution: at $x$ below the average
	the environment feeds the chain, above it detracts from the system. 
	Since $\exv{\gamma}_Q=0$, there is no extra feed at the beginning of the chain.
}

\end{figure}

A similar arrangement of the rates in the continuous model,
\be
 \gamma(x) \: = \: \sigma (ax-c),
 \qquad 
 \mu(x) \: = \: \sigma x, 
\ee{LINGAMMU}
leads to the {\em two-parameter gamma distribution},
\be
 Q(x) \: = \: \frac{K}{\sigma x} \, \ead{-\int (a-c/x) \, dx} \: = \: 
 \frac{a^c}{\Gamma(c)} \, x^{c-1} \, \ead{-ax}.
\ee{GAMMAFROMLIN}
This stationary distribution emerges as a result
of a pure (unthresholded) linear preference in the growth rate and a linear, but not
overall positive loss rate to the environment. 
The negative values of $\gamma(x)$
actually mean a feeding from the environment (Fig.\ref{GAMMAFIG}).

Such a gamma distribution fits income data very well \cite{JAKOVENKO}. 
We risk the conclusion
that in the background of such processes, beyond the linear prefrence
rate, $\mu(x)=\sigma x$ (often cited as the Matthias principle in market
economies), a taxation and a social welfare redistribution system acts. 







\section{Evolution of entropic distance}

The entropy -- probability connection is also interpreted as a measure of a distance to the 
minimal information state. The well-known Boltzmann--Gibbs--Shannon formula
is a special instance of the more general entropic distance, $\rho(P,\Pi)$ between
two distributions. For such generalized entropic distances 
the following requirement should hold: 
$\rho(P,\Pi) \ge 0$, and reaches zero only for identical distributions ($\rho(P,P)=0$
and $\rho(P,\Pi) > 0$ for $P \ne \Pi$). We consider in this paper univariate distributions,
$P_n$, $\Pi_n$ indexed with a natural number, $n$, and normalized as $\sumi P_n=1$
and $\sumi \Pi_n = 1$, respectively. 

By considering random dynamics in fast growing complex systems, dominantly unidirectional
changes in the quantity $n$ are considered. The general question arises that whether
there exist a quantity, possibly expressed as an expectation value of a function of
the respective probability values at the same state indexed by $n$, which changes only
in one direction during the dynamical evolution. In particular the entropic distance
to a stationary distribution, $Q_n$, from any starting distribution, $P_n=P_n(0)$,
should decrease during such an evolution: 
\be
 \pt{}{t} \, \rho(P,Q) \: \le \: 0.
\ee{RHOHASTODECREASE}
A trace form for the entropic distance from a non-constant stationary $Q_n$ is given by
\be
 \rho(P,Q) \: = \: \sumi \, \mathfrak{s} (P_n,Q_n) \, Q_n.
\ee{ENTROPICDISTANCE}
It is very common to deal with entropic distances defined via a function of
the ratio of the respective probabilities only, $\mathfrak{s}(\xi_n)$ with $\xi_n=P_n/Q_n$.
Then, from the property of zero distance from itself one concludes that
$\mathfrak{s}(1)=0$, and it is different from zero if there is an index $n$
such that $\xi_n \ne 1$.


The change of the entropic distance is governed by its definition and the evolution 
equation for the distribution.
The entropic distance of an actual, time-dependent distribution, $P_n(t)$, to the
stationary distribution, $Q_n$ has the trace form \cite{BIROSCHRAM}:
\be
 \rho \: = \: \sum_n\limits \mathfrak{s}(\xi_n) Q_n.
\ee{RHODEF}
For all concave $\mathfrak{s}(\xi)$ functions the following Jensen inequality applies:
\be
 \rho \: \ge \: \mathfrak{s}\left(\sum_n \xi_n Q_n \right) =
 \mathfrak{s}\left(\sum_n P_n \right) = \mathfrak{s}(1) = 0.
\ee{RHODEFJENSEN}
The time derivative of thi entropic distance is given by
\be
 \dot{\rho} \: = \: \sum_n\limits \mathfrak{s}^{\prime}(\xi_n) \dot{\xi}_n Q_n
 \: = \: \sum_n \mathfrak{s}^{\prime}(\xi_n) \dot{P}_n.
\ee{RHODOT0}
Now we utilize eq.(\ref{GENMARKOV}) and obtain
\be
 \dot{\rho} \: = \: \sum_{n,m}\limits \mathfrak{s}^{\prime}(\xi_n) \, 
 \left[w_{n\leftarrow m}Q_m\xi_m - w_{n\leftarrow m} Q_n\xi_n \right].
\ee{RHODTOMARK1}
As a first step we write $\xi_m = \xi_n + (\xi_m-\xi_n)$ and use the property
\be
 0 \: = \: \sum_m\limits \left[ w_{n\leftarrow m}Q_m \, - \, w_{m\leftarrow n} Q_n \right]
\ee{GENMARKOVSTAT}
for the stationary state. The above formula transforms to
\be
 \dot{\rho} \: = \: \sum_{n,m}\limits \mathfrak{s}^{\prime}(\xi_n) \,
 (\xi_m - \xi_n) w_{n\leftarrow m} Q_m.
\ee{RHODOTMARK2}
In the second step we use the remainder theorem for Taylor series in its Lagrange form:
\be
 \mathfrak{s}(\xi_m) \: = \: \mathfrak{s}(\xi_n) + (\xi_m - \xi_n) \mathfrak{s}^{\prime}(\xi_n)
 + \frac{1}{2} (\xi_m-\xi_n)^2 \mathfrak{s}^{\prime\prime}(c_{nm})
\ee{LAGRANGE}
with the internal point $c_{nm}$ lying between $\xi_n$ and $\xi_m$.
Expressing the first order term in eq.(\ref{LAGRANGE}), eq.(\ref{RHODOTMARK2}) becomes
\ba
 \dot{\rho} \: &=& \: \sum_{m,n}\limits \left[\mathfrak{s}(\xi_m)-\mathfrak{s}(\xi_n) \right]
 w_{n\leftarrow m}Q_m \, - \, 
\nl
 \, &-& \, \frac{1}{2} \sum_{n,m}\limits (\xi_m-\xi_n)^2 \mathfrak{s}^{\prime\prime}(c_{nm}) \,
 w_{n\leftarrow m}Q_m. 
\ea{RHODOTMARK3}
Here the first sum on the right hand side vanishes due to the stationarity
(\ref{GENMARKOVSTAT}). This can be seen by exchanging the summation indices $m$ and $n$ in the first
part, leading to
\be
 \sum_{n}\limits \mathfrak{s}(\xi_n) \sum_m\limits 
 \left[w_{m\leftarrow n} Q_n - w_{n\leftarrow m}Q_m \right] \: = \: 0.
\ee{RHODOTMARKVANISH}
For positive transition rates, $w_{n\leftarrow m} > 0$, the remainder term
is always negative for concave, $\mathfrak{s}^{\prime\prime}(\xi) > 0$ functions.

In the special case of the avalanche dynamics with loss,
$w_{n\leftarrow m} = \mu_m \delta_{m,n-1} + \gamma_m \delta_{n,0}$, 
we obtain
\ba
 \dot{\rho} \: &=& \: 
 - \frac{1}{2} \sum_{n}\limits (\xi_n - \xi_{n-1})^2 \mathfrak{s}^{\prime\prime}(c_{n,n-1}) \mu_{n-1} Q_{n-1}
\nl
\, &-& \,  \frac{1}{2} \sum_{n}\limits (\xi_n - \xi_0)^2 \mathfrak{s}^{\prime\prime}(c_{n,0}) \gamma_n Q_n.
\ea{SPECRHODOT}
For positive rates $\gamma_n$ and $\mu_n$ therefore $\dot{\rho} < 0$ unless the equilibrium
state is achieved where $\xi_n=1$ for all $n$.


Finally let us briefly discuss cases when some $\gamma_n$ can be negative.
We encountered this for processes leading to negative binomial or gamma distributions.
The remainder result (\ref{SPECRHODOT}) in such a case does not guarantee
a steady approach towards the stationary distribution  in terms of a general
entropic distance. 
Henceforth further investigations are necessary.

\section{Conclusion}

In the present work we have proposed a unified mathematical framework 
based on a master equation approach to complex systems governed by random
dynamics. In particular we have focused on transition rates which do not lead
to a detailed balance. A wide variety of stationary distributions known from
complex network research, particle physics, scientometrics, econophysics,
biology and demography are successfully reproduced.

This view is able to clarify why only the linear preference rate leads to
a power-law tailed degree distribution in random
networks as well as to a Pareto-type distribution of wealth when the 
preference expressed by ''the rich gets richer'' principle is linear. 
Similarly, the distribution of scientific citations, known to be power-law tailed, has been
explained earlier on the basis of such an evolution equation\cite{SCHUBERT}.
The exponential (geometrical) distribution is obtained for constant rates and
the power-law tailed Waring (in the continuum limit Tsallis--Pareto) distribution
for a linear pereference growth rate. The method outlined in this paper is able to
deliver further well-known and frequently used distributions, such as the
Weibull or the Gompertz distribution or the stretched exponential. 

Beyond the above mentioned practical application possibilities
we have established connections to the fundamental fluctuation--dissipation
relation central in statistical physics.
In the simplified version with a constant loss rate, $\gamma_n=\gamma$,
the stationary PDF, $Q(x)$, 
are proved to be related to quantities familiar from
general statistics: the necessary growth rate is reciprocial to the
hazard rate, $\mu(x)=\gamma/h(x)$. The correspondence between this hazard rate
and the cumulated hazard was generalized to a ''fluctuation--dissipation''
type relation between the growth rate, $\mu(x)$, and the loss rate $\gamma(x)$ in 
eq.(\ref{FLUDICONT}). 
A similar general relation was derived for the
discrete version in eq.(\ref{MUZERO}). The specific case $m=0$
gives the key to reconstruct
the first attachment rate $\mu_0$ from observing $Q_n$ and measuring  
$\exv{\gamma}_Q$:
$ \mu_0 \: = \: \exv{\gamma}_Q / Q_0 \, - \, \gamma_0$.

Finally, while seeking answer to the question which entropy formula could be the
optimal one for such unbalanced growth processes in random systems,
we proved that any entropic distance based on a general concave function
of the probability ratio, $\mathfrak{s}(\xi)$, will decrease  to zero for $\gamma(x) > 0$.

Generalizing further the dynamics for $\gamma_n$ containing both positive and negative
elements we have discussed two models. First with $\gamma_n=\sigma(n-fk)$
and $\mu_n=\sigma f(n+k)$ a model for high energy hadron production, leading to
a negative binomial stationary distribution was evoked. Second, with $\gamma(x)=\sigma (ax-b)$
and $\mu(x)=\sigma x$ a continuous model for the income distribution was recited.
This model assumes a constant percentage taxation and social welfare amandments, leading to
a gamma distribution. 


The unified mathematical treatment outlined in this paper should be a primary tool
in understanding intriguing universality classes reported in complex systems.
Important questions are left open for further research: what are the precise
conditions for entropy growth in cases involving partially negative $\gamma(x)$ rates
(while $\gamma(x)+\mu(x)>0$ is still satisfied); 
what are the minimal conditions for gaining a stationary distribution in
unbalanced random processes;
or how the transient dynamics towards the stationary state is displayed with time.

\section*{Acknowledgement}

This work has been supported by the Hungarian Scientific Research Fund OTKA,
supervised by the National Research, Development and Innovation Office NKFIH
(pro\-ject No.104260) and by a UBB STAR fellowship.
Discussions with A.~Telcs and Zs.~L\'az\'ar are gratefully acknoweldged. 
Z.~N\'eda acknowledges support from PN-II-ID-PCE-2011-3-0348 research grant.



\section*{Appendix}

Here we prove eq.(\ref{CONSISTENT}). We define
the summed expression of product chain as
\be
 S_0 \: = \: \sum_{n=0}^{\infty}\limits r_n \, \prod_{i=0}^n \frac{1}{1+r_i}.
\ee{SZERO}
The first few terms are:
\be
 S_0 \: = \: \frac{r_0}{1+r_0} + \frac{1}{1+r_0} \frac{r_1}{1+r_1} +
 \frac{1}{1+r_0} \frac{1}{1+r_1} \frac{r_2}{1+r_2} + \ldots
\ee{SZERO3}
By rearranging the sum starting at the second term,
\be
 S_0 \: = \: \frac{r_0}{1+r_0} + \frac{1}{1+r_0} \left( \frac{r_1}{1+r_1} +
  \frac{1}{1+r_1} \frac{r_2}{1+r_2} + \ldots \right),
\ee{SZEROFAC}
we realize that
\be
 S_0 \: = \: \frac{r_0}{1+r_0} \, + \, \frac{1}{1+r_0} S_1
\ee{SZERONE}
with an obviuos notation, $S_1$, for the same infinite sum starting with terms containing $r_1$.
After a linear re-arrangement it is convincing that this relation,
\be
 \left( S_0 - S_1 \right) \: = \: r_0 \left( 1 - S_0\right),
\ee{SZEROFINAL}
holds for an arbitrary $r_0$ if and only if $S_1=S_0=1$.
The same proof is valid for starting at any $m$-th element. 
$S_0=1$ proves the original statement.

\end{document}